\documentclass{aastex61}
\usepackage{amsmath}

\received{}
\revised{}
\accepted{}
\submitjournal{ApJ}

\begin{document}

\title{On the bimodal spin period distribution of Be/X-ray pulsars}

\correspondingauthor{Xiang-Dong Li}
\email{lixd@nju.edu.cn}

\author{Xiao-Tian Xu}
\affil{School of Astronomy and Space Science, Nanjing University, Nanjing 210046, China}
\affil{
Key Laboratory of Modern Astronomy and Astrophysics (Nanjing University), Ministry of Education, Nanjing 210046, China
}
\author{Xiang-Dong Li}
\affil{School of Astronomy and Space Science, Nanjing University, Nanjing 210046, China}
\affil{
Key Laboratory of Modern Astronomy and Astrophysics (Nanjing University), Ministry of Education, Nanjing 210046, China
}

\begin{abstract}
It has been reported that there are two populations of Be/X-ray pulsars, with the pulse period distribution peaked at
$\sim 10$ s and $\sim 200$ s, respectively. A possible explanation of this bimodal distribution is related to different accretion modes in Be/X-ray binaries. In this work, we investigate the spin evolution of Be/X-ray pulsars based on the magnetically threaded accretion disk model. Compared with previous works, we take into account several distinct and important factors of Be/X-ray binaries, including the transient accretion behavior and possible change of the accretion disk structure during quiescence. We demonstrate that current Be/X-ray pulsars are close to the spin equilibrium determined by the balance of spin-up during outbursts and spin-down during quiescence, and that the observed bimodal distribution can be well reproduced by the equilibrium spin periods with reasonable input parameters.
\end{abstract}


\section{Introduction} \label{sec:intro}
Be/X-ray binaries (BeXBs) are a sub-class of high-mass X-ray binaries, usually composed of a neutron star (NS) and an O/Be type star \citep{Rei2011}. The O/Be stars in BeXBs are generally fast-rotating main-sequence stars with a circumstellar disk, featured by Balmer emission lines.  These observational properties can be well reproduced by the decretion disk model \citep{Lee1991},  which is similar to an accretion disk except the inverse direction of mass flow. Note that the circumstellar disk is not a stable structure, undergoing dissipation and recurrence \citep{Hai1999}, the size and the structure of which are strongly affected by the gravity and the orbital motion of the NS \citep{Neg2001a,Neg2001b, Oka2002}. Part of the disk material is captured by the NS, preferentially near periastron. The accreted material is then channeled onto the NS surface by its magnetic field lines, so the NS usually appears as an Be/X-ray pulsar (BeXP) \citep{Lam1973}.

BeXBs are further divided into persistent and transient sources. A small population of BeXBs are persistent sources, characterized by nearly circular orbits (with eccentricities $e \lesssim 0.2$) and low X-ray luminosities ($L_{\rm X}\sim 10^{34}-10^{35}$ erg\,s$^{-1}$), while other BeXBs are transient sources with eccentric orbits (with $e \gtrsim 0.3$)  \citep{Rei2011}. They experience outbursts from time to time, which are separated by long, quiescent intervals. There are two types of outbursts in transient BeXBs. Type I or normal outbursts are characterized by (quasi-)periodicity with peak luminosities ranging from $\sim 10^{36}$ to $\sim 10^{37}$ erg\,s$^{-1}$. They are thought to be caused by rapid accretion of the NS when it passes periastron. Type II or giant outbursts show non-periodicity, and usually have peak luminosities $\gtrsim 10^{37}$ erg\,s$^{-1}$. The origin of type II outbursts is still open to debate.

\citet{Kni2011} found that the distribution of the spin periods ($P$) of BeXPs in the Milky Way (MW) and the Small/Large Magellanic Clouds (SMC/LMC) show a bimodal feature, peaked at
$\sim 10$ s and $\sim 200$ s, respectively. These authors proposed that the two populations may be related to two types of supernovae, i.e., iron core-collapse supernovae and electron-capture supernovae, respectively. \citet{Che2014} suggested that the two populations of BeXBs are likely associated with different accretion modes in the two types of outbursts. Because the peak luminosity and the accretion disk structure are different in type I and II outbursts, NSs undergoing different types of outbursts are subject to different accretion torques, which lead to bimodal distribution of the spin periods.

The accretion process in BeXBs is very complicated and uncertain. Although it is generally believed that the NS undergoes disk accretion during outbursts, scenarios involving wind accretion have also been proposed \citep{Ikh2007,Sha2012}. Even for disk accretion, there were arguments that  the viscous timescale of the standard thin disk \citep{Sha1973} fails to explain the observed outburst duration of type I outbursts, suggesting an advection-dominated accretion flow (ADAF) instead \citep{Oka2013}.

Several authors \citep{Cha2012,Klu2014,Shi2015} investigated the long-term spin evolution of BeXPs in the SMC, and arrived the conclusion that a large population of the BeXPs are magnetars, i.e., NSs with surface magnetic fields stronger than $4\times 10^{13}$ G. But they are in conflict with the observations of cyclotron lines in the Galactic X-ray pulsars \citep[][and references therein]{Cob2002, Cab2012,Wal2015}.

Considering the fact that most BeXBs are transient sources, the transition between outbursts and quiescence should play an important role in the spin evolution of the NSs. In most of the previous studies, only the long-term average evolution was taken into account, which actually deviates from the evolution in the case of transient accretion.
In this work, we attempt to investigate the spin evolution in BeXPs in a more self-consistent way by considering accretion in both outbursting and quiescent stages. We take into account the possible influence of various types of outbursts and the structure of the accretion flows.  In section 2, we describe our theoretical considerations. In section 3, we describe the data sample and obtain possible constraints on the model parameters through comparison with observations. We summarize our work  in section 4.

\section{Model}

\subsection{Spin evolution}
We assume that the NS in a BeXB is interacting with an accretion disk originating from the Be star's wind. We first introduce the light-cylinder radius $R_{\rm L}$ and the inner radius of the accretion disk $R_{\rm 0}$. Here, $R_{\rm L}$ is defined to be
\begin{equation}
R_{\rm L}=\frac{c}{\Omega_{\rm S}},
\end{equation}
where $c$ is the speed of light and $\Omega_{\rm S}$ is the angular velocity of NS;
$R_0$ is evaluated by
\begin{equation}
R_0=\phi R_{\rm A},
\end{equation}
where $\phi$ is a factor around unity \citep{Gho1979a,Gho1979b,Lon2005}, taken to be 0.5 in our work, and $R_{\rm A}$ is the Alfv\'en radius for spherical accretion
\begin{equation}
R_{\rm A}=\left(\frac{\mu^2}{\dot{M}\sqrt{2GM}}\right)^{2/7},
\end{equation}
where $\mu$ is the magnetic moment of the NS, $\dot{M}$  the accretion rate, $G$  the gravitational constant, and $M$ the mass of the NS, taken to be $1.4~M_\sun$.
According to the relation between $R_{\rm L}$ and $R_0$, the spin evolution is divided into two states: rotation-powered state ($R_{\rm L} < R_{\rm 0}$) and accretion-powered state ($R_{\rm L} > R_0$)\footnote{We combine both the accretor and propeller states into the accretion-powered state since mass ejection during the propeller phase also requires extraction of accretion power.}.

In the rotation-powered state, the spin evolution of the NS is determined by the torque $N_{\rm dip}$ due to magnetic dipole radiation \citep{Sha1983}, i.e.
\begin{equation}
N_{\rm dip}=-\frac{\mu^2\Omega_{\rm S}^3}{9c^3}.
\end{equation}
Considering the condition $R_{\rm L}=R_0$, we obtain the critical spin period that separates the rotation- and accretion-powered states,
\begin{equation}
P_{\rm cr}\approx 0.23 \left(\frac{\mu}{10^{30}{\rm ~G~cm^{3}}}\right)^{4/7}
\left(\frac{\dot{M}}{10^{14}{\rm ~g~s^{-1}}}\right)^{-2/7} {\rm ~s.}
\end{equation}
The critical spin period is much smaller than the current spin periods of most BeXPs with typical magnetic fields and accretion rates.
Therefore, the rotation-powered state is usually ignorable.

In the accretion-powered state, we assume that the spin evolution of the NS is dominated by disk accretion and the interaction between the disk and the NS magnetic field, according to the magnetically threaded accretion disk model \citep{Gho1979a,Gho1979b}. The accretion torque $N_{\rm d}$ exerted by the disk consists of two terms,
\begin{equation}
N_{\rm d}=N_0+N_{\rm M}.
\label{N1}
\end{equation}
Here $N_0$ is the material torque due to mass accretion,
\begin{equation}
N_0=\dot{M}\Omega_{\rm K}(R_0)R_0^2,
\end{equation}
where $\Omega_{\rm K}$ is the Keplerian angular velocity in the disk, and $N_{\rm M}$ is the torque resulting from the magnetic field-disk interaction, due to the shearing motion between the magnetic field lines of the NS and the disk material.
Equation~(\ref{N1}) can be expressed in the following form \citep{Gho1979a,Gho1979b},
\begin{equation}
N_{\rm d}=N_0n(\omega),
\label{N2}
\end{equation}
where $\omega$ is the fastness parameter defined by
\begin{equation}
\omega\equiv\frac{\Omega_{\rm S}}{\Omega_{\rm K}(R_0)},
\end{equation}
and $n(\omega)$ is a dimensionless function, which is, according to \citet{Gho1979a,Gho1979b},
\begin{equation}
n\approx 1.39\{1-\omega[4.03(1-\omega)^{0.173}-0.878]\}(1-\omega)^{-1}.
\label{GL1}
\end{equation}
This equation was derived based on the assumption that the NS magnetic field penetrates the disk via the Kelvin-Helmholtz instability, turbulent diffusion, and reconnection with small-scale fields within the disk. However, \citet{Wang1987} pointed out that this will lead to the pressure of the wound field to exceed the thermal pressure at large radius and disrupt the disk. There are various forms of the $n(\omega)$ function proposed in the literature \citep[e.g.,][]{Wan1995,Li1996,KR2007}. Here we adopt the following simplified equation
\begin{equation}
n\approx 1-\frac{\omega}{\omega_{\rm c}},
\label{GL2}
\end{equation}
which is similar to the torque equation in \citet{Lip1982} and used in previous studies \citep[e.g.][]{Li1999,Har2002,Ho2014,Eks2015}. In Eq.~(\ref{GL2}), $\omega_{\rm c}$ is a critical value of the fastness parameter at which $n=0$. This means that $N_{\rm d}>0$ when $\omega<\omega_{\rm c}$ and $N_{\rm d}<0$ when $\omega>\omega_{\rm c}$.  Also note that Eq.~(\ref{GL2}) can be used to cover both accretor and propeller states: when $\omega\rightarrow 0$, $N_{\rm d}=\dot{M}R_0^2\Omega_{\rm K}(R_0)$; when $\omega\gg 1$, $N_{\rm d}\sim-\dot{M}R_0^2\Omega_{\rm S}$.

It should be pointed out that Eq.~(\ref{N2}) is only applicable to steady accretion onto NSs, while the accretion processes in BeXBs are much more complicated, so we consider further modifications of the accretion torque by taking into the following factors.

(1) Due to the shear between the magnetosphere of the NS and the accretion disk, there is an open-field region on the magnetosphere, where the accreting material can potentially escape as wind \citep{Lov1995,Rom2003}. The coupling between the magnetic field of the NS and the wind material generates a spin-down torque on the NS. In our model, we assume  that a fraction $\eta$ of the accreting material leaves the system as wind, i.e., $\dot{M}_{\rm w}=\eta \dot{M}$. Then, the torque exerted by the wind $N_{\rm w}$ is given by
\begin{equation}
N_{\rm w}=-\eta_{\rm w}\dot{M}_{\rm w}\sqrt{GMR_{\rm A,w}},
\end{equation}
where $R_{\rm A,w}$ is the Alfv\'en radius of the wind, and $\eta_{\rm w}$ is a efficiency coefficient for the magnetic field-wind interaction. In our work, we take $\eta_{\rm w}=1$. Then, the total torque becomes
\begin{equation}
N=N_{\rm d}+N_{\rm w}=N_{\rm 0}n(\omega)+N_{\rm w}.
\label{n0}
\end{equation}

(2) Most BeXBs are transient sources subject to (quasi-)periodic type I outbursts \citep{Rei2011}.
The spin evolution during type I outbursts is described by
\begin{equation}
N_{\rm I}=N_{\rm 0,I}n(\omega_{\rm I})+N_{\rm w,I},
\end{equation}
where the subscripts I represents the parameters evaluated for type I outbursts.
Assuming that the detection of X-ray pulsations marks the occurrence of the outburst phase, we derive that the duty cycle $x$  of type I outbursts roughly ranges from 0.005 to 0.05 from the data sample compiled by \citet{Yan2017} .

(3) Type II outbursts are rare but more violent compared with type I outbursts. The origin and accretion physics of type II outbursts are still open to debate.  While \citet{Oka2013} argued that the Bondi-Hoyle-Lyttleton accretion is responsible for type II outbursts, disk accretion seems to be observationally preferred \citep[e.g.,][]{Hai1999,Sug2017}. Here we adopt the suggestion by \citet{Sug2017} and
write the torque $N_{\rm II}$ during type II outbursts to be
\begin{equation}
N_{\rm II}=N_{\rm 0,II}n(\omega_{\rm II})+N_{\rm w,II},
\end{equation}
where the subscript II represents that the parameters are evaluated for type II outbursts.
We introduce the duty cycle $y$ to describe the occurrence of type II outburst,  given by
\begin{equation}
y={\rm outburst~rate}\times{\rm outburst~duration}.
\end{equation}
The observations by \citet{Sug2017} suggest that $y$ is about 0.1 with the outburst rate $\sim 1/1000 {\rm ~day}^{-1}$ and the outburst duration $\sim 100 {\rm ~days}$.

(4) During the quiescent phase BeXBs are very faint with X-ray luminosities $L_{\rm X,l}$ as low as $\sim10^{33}{\rm ~erg~s^{-1}}$ \citep{Yan2017}, so the accretion disk may evolve from an optically thick, geometrically thin disk to an ADAF. Consequently, the accretion disk becomes geometrically thick with sub-Keplerian rotation, so we add a parameter $A$ to modify the disk rotation for an ADAF,
\begin{equation}
\Omega_{\rm D}=A\Omega_{\rm K}.
\end{equation}
The typical value of $A$ is $0.2-0.3$ \citep{Nar1995}. Then,the torque  $N_{\rm q}$ in the quiescent phase can be rewritten to be
\begin{equation}
N_{\rm q}=AN_{\rm 0,q}n(\omega_{\rm q}/A)+N_{\rm w,q},
\label{Nq}
\end{equation}
where the subscript q represents the parameters evaluated in the quiescent phases.

With the above factors taken into account, the {\em averaged} total torque exerted by an accretion disk is written to be
\begin{equation}
N\approx xN_{\rm I}+ yN_{\rm II}+(1-x-y)N_{\rm q}.
\label{Ntot}
\end{equation}
Combining Eqs.~(\ref{N1})-(\ref{Ntot}) and setting $N=0$, we obtain an approximate estimate of the equilibrium spin period $P_{\rm eq}$ for the NS in a transient BeXB (see Appendix for the derivation),
\begin{equation}
P_{\rm eq}\simeq
\left(\frac{2\pi\phi^{3/2}}{2^{3/14}}\right)
\left[\frac{\mu^{6/7}}{\omega_{\rm c}(GM)^{5/7}}\right]
\left[\frac{x\dot{M}^{3/7}_{\rm I}+(1-x-y)\dot{M}^{3/7}_{\rm q}+y\dot{M}_{\rm II}^{3/7}}
{x\dot{M}^{6/7}_{\rm I}C_{1,1}^{-1}+(1-x-y)\dot{M}^{6/7}_{\rm q}C_{\rm 1,A}^{-1}+
y\dot{M}^{6/7}_{\rm II}C_{1,1}^{-1}}\right],
\label{peqGL}
\end{equation}
where
\begin{equation}
C_{1,1}=\left(1-\frac{\eta^{6/7}}{\phi^{1/2}}\right)^{-1},{~\rm and~}C_{1,A}=\left(A-\frac{\eta^{6/7}}{\phi^{1/2}}\right)^{-1}.
\end{equation}
In addition, we require that the magnitude of the spin-down torque generated by the wind is always weaker than that of the material torque, that is $C_{1,1}{\rm~and~}C_{\rm 1,A}>0$.

\subsection{Spin-change timescale and spin equilibrium}
In the accretion-powered state, the spin evolution of the NS is determined by the total torque exerted on the NS by the accretion disk
\begin{equation}
-2\pi I\frac{\dot{P}}{P^2}=N,
\label{Nns}
\end{equation}
where $I$ is the moment of inertia of the NS and $\dot{P}$ is the derivative of the spin period.
This equation can be solved analytically, and according to its solution, we obtain an estimate of the typical spin-change timescale $\tau_{\rm spin}$ (see Appendix for the derivation),
\begin{equation}
\begin{split}
\tau_{\rm spin}
=\left[\frac{I\omega_{\rm c}}{\phi^2}\frac{(2GM)^{2/7}}{\mu^{8/7}}\right]
\left(\frac{GM}{\bar{L}_{\rm X}R}\right)^{3/7},
\label{tauGL}
\end{split}
\end{equation}
where $\dot{M}$ is converted into $L_{\rm X}$ via
\begin{equation}
L_{\rm X}=\frac{GM\dot{M}}{R},
\end{equation}
$R$ is the radius of the NS, and $\bar{L}_{\rm X}$ is the averaged X-ray luminosity, defined by
\begin{equation}
\bar{L}_{\rm X}^{3/7}=xL_{\rm X,I}^{3/7}+(1-x-y)L_{\rm X,q}^{3/7}+yL_{\rm X,II}^{3/7}.
\end{equation}

It is interesting to note that $\tau_{\rm spin}$ is independent on the structure of the accretion disk and the wind loss, though these two factors affect the value of $P_{\rm eq}$.
Eq.~(\ref{tauGL}) shows that $\tau_{\rm spin}\propto \bar{L}_{\rm X}^{-3/7}\mu^{-8/7}$, i.e., systems with higher average accretion rate and stronger magnetic field can reach spin equilibrium with shorter time.
Assuming steady accretion and adopting the following parameters: $x=0$, $y=0$, $L_{\rm X,q}=10^{34}$ erg s$^{-1}$, $\mu=10^{30}$ G\,cm$^3$, $R=10^{6}$ cm, $M=1.4~M_\sun$, $I=10^{45}$ g\,cm$^2$, $\phi=0.5$, and $\omega_{\rm c}=0.8$, we obtain $\tau_{\rm spin}\lesssim 10^6$ yr, which implies that a typical NS is able to reach the spin equilibrium within $10^6{\rm ~ yr}$ even if we only take into account accretion during quiescence. Obviously $\tau_{\rm spin}$ become shorter if we consider the contribution from type I/II outbursts.
Considering the ages of most BeXBs are roughly 10 Myr \citep{Ant2010}, it is safe to assume that most of observed BeXPs have evolved to be near the spin equilibrium.

\begin{figure}
\centering
\plotone{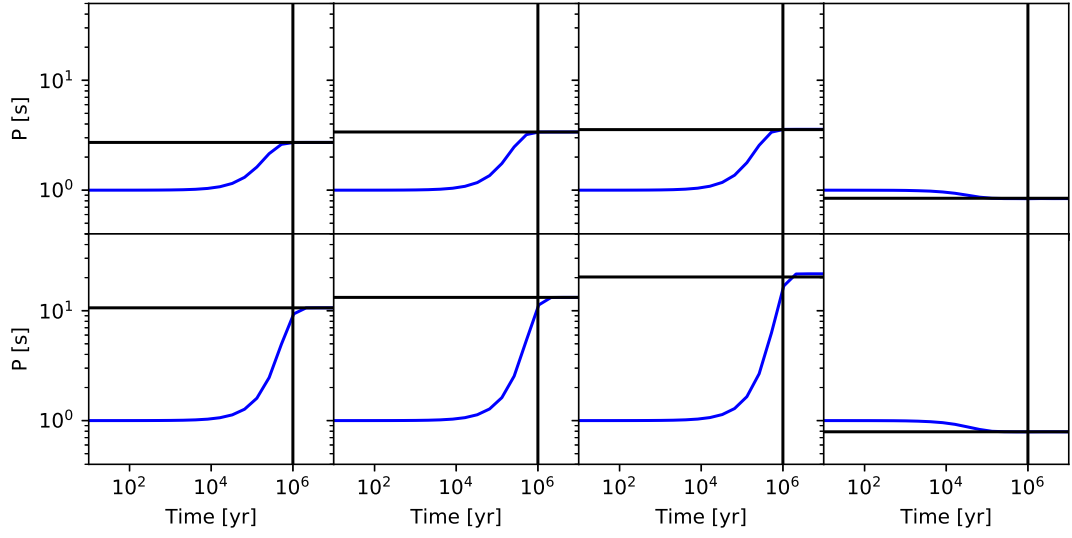}
\caption{
The calculated spin period evolution in BeXPs. The upper and bottom panels show the results with x = 0.05 and
0.005, respectively. In the first, second, third, and fourth columns, we set
$(A,~y,~\eta)$ to be (1, 0, 0), (1, 0, 0.1), (0.1, 0, 0.1), and
(0.1, 0.1, 0.1), respectively.
The blue line, black vertical line, and black horizontal line represent the evolutionary tracks of $P$,
the age of $10^6$ yr, and the equilibrium spin period $P_{\rm eq}$ given by Eq. (\ref{peqGL}), respectively.
}
\label{f1}
\end{figure}

We then perform a numerical test of the above analysis. We adopt the following parameters:
$P_0=1$ s, $(L_{\rm X,I},~L_{\rm II},~L_{\rm X,q})=(10^{37},~10^{38},~10^{34}){\rm ~erg~s^{-1}}$, $\mu=10^{30}$ G\,cm$^3$, $R=10^6$ cm, $A=0.1$ or 1, $x=0.05$ or 0.005, $y=0$ or 0.1, $P_{\rm orb}=100$ days, $\eta=0$ or 0.1,
and $\omega_{\rm c}=0.8$.  Our results are shown in Fig.~\ref{f1}. In this figure, the upper and lower panels depict the results with $x=0.05$ and $0.005$, respectively. In the first, second, third, and fourth columns, we set $(A,~y,~\eta)$ to be $(1,~0,~0)$, $(1,~0,~0.1)$, $(0.1,~0,~0.1)$, and $(0.1,~0.1,~0.1)$, respectively. In each panel, the blue line represents the evolutionary track of $P$, the black vertical line the age of $10^{6}$ yr, and the black horizontal line the equilibrium spin period $P_{\rm eq}$ given by Eq.~(\ref{peqGL}).
The numerical calculations confirm our analytical result that current BeXPs should be close to spin equilibrium.
A remarkable feature is that type II outbursts can significantly change the overall distribution of the equilibrium spin periods.
\citet{Che2014} showed that the BeXPs with $P<40$ s are more likely to experience type II outbursts compared with those with $P>40$ s. The calculated results presented in Fig.~\ref{f1} demonstrates that the NSs can be significantly spun up during type II outbursts.

In next section we perform statistical calculations to examine whether the equilibrium spin periods in transient accretion can reproduce the bimodal distribution of BeXPs.

\section{The bimodal distribution}
\begin{figure}
\plotone{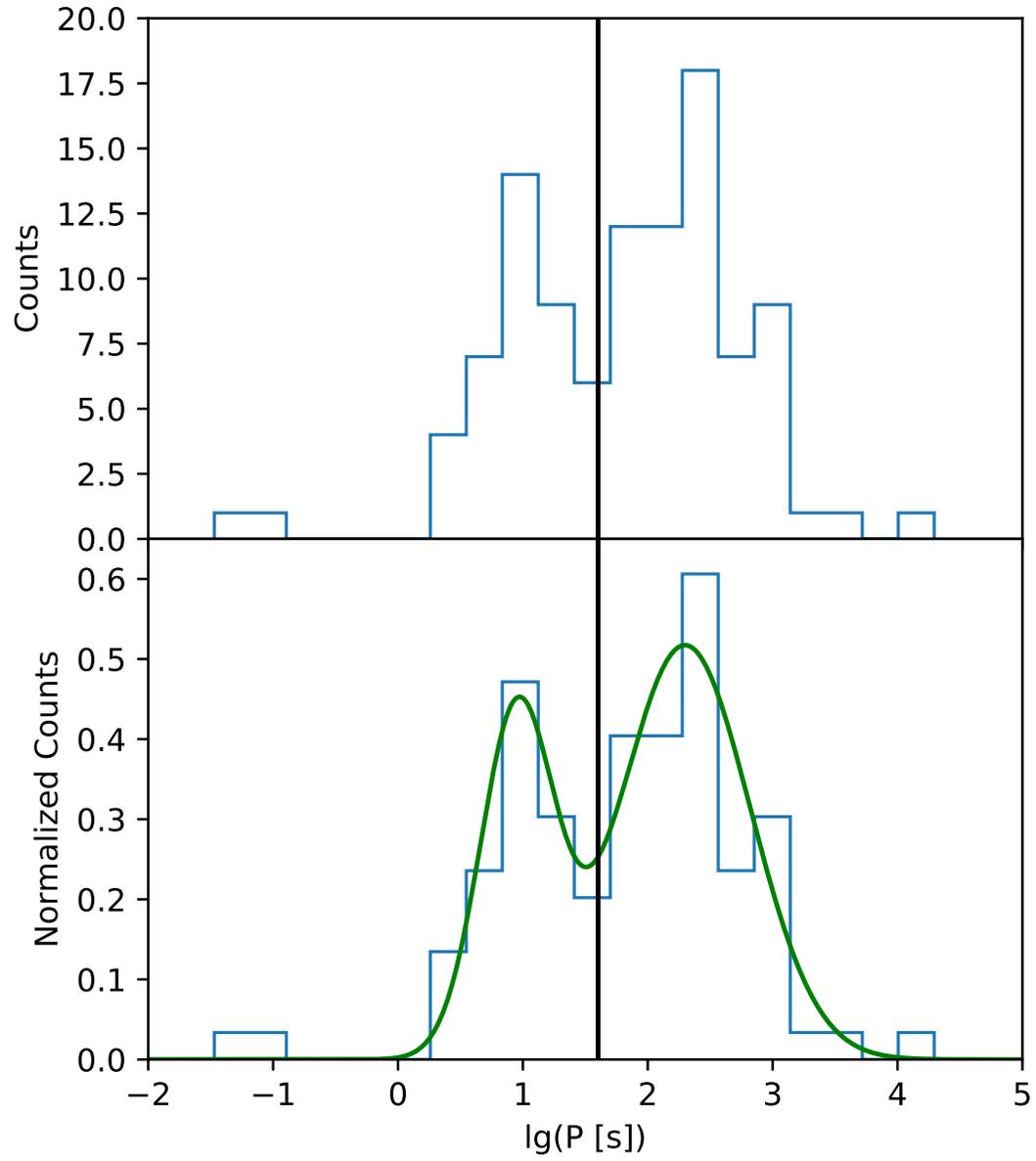}
\caption{
The distribution of the spin periods of NSs in 103 BeXBs. The upper panel shows the observed distribution of $P$. The lower panel shows the normalized distribution of $P$ (blue solid line) and the fitting curve with a double-Gaussian function (green solid line). The black vertical line represents $P=40{\rm ~s}$.
\label{f2}}
\end{figure}

We first describe the data sample that will be used to compare with the model prediction. There are 103 BeXPs in the sample, including the sources from MW, SMC, and LMC  \citep[data taken from][]{Rag2005,Rei2011,Tow2011,Stu2012b,Vas2014,Coe2015,Vas2016,Vas2017,Yan2017}. The distribution of the pulse period $P$ is presented in Fig.~\ref{f2}, where the black vertical line represents $P=40$ s. The upper panel shows the histogram of the real $P$ distribution, and the lower panel depicts the normalized distribution of $P$ (blue solid line) and the fitted curve (green solid line)\footnote{ The fitting process was performed with the Scipy library \citep{Eri2001}.} with a double-Gaussian function, which is defined as
\begin{equation}
F(x)=\frac{A}{\sqrt{2\pi \sigma_1^2}}{\rm exp}\left[-\frac{(x-\mu_1)^2}{2\sigma_1^2}\right]+
\frac{1-A}{\sqrt{2\pi \sigma_2^2}}{\rm exp}\left[-\frac{(x-\mu_2)^2}{2\sigma_2^2}\right].
\end{equation}
The fitting parameters are as follows\\
$A=0.321\pm0.057$;\\
$\mu_1=0.953\pm0.056$, $\mu_2=2.298\pm0.067$;\\
$\sigma_1=0.296\pm0.063$, $\sigma_2=0.524\pm0.071$,\\
where $x$, $\mu_{1,2}$, and $\sigma_{1,2}$ are in the units of second, and the error for each parameter was calculated by the fitting method\footnote{See https://docs.scipy.org/doc/scipy/reference/generated/scipy.optimize.curve\_fit.html for details.}. We apply the Hartigans' dip test \citep{Har1985} to our sample. The P-value is 0.4308 for the distribution in logarithmic scale, and 0.9918 in linear scale\footnote{The Hartigans' dip test was performed with the R based library diptest \citep{Mar2016, R2018}.}

Then, we introduce the properties of the input parameters for our statistical calculation,
\begin{itemize}
\item The duty cycle $x$ of type I outbursts: based on the data compiled in \citet{Yan2017} for the occurrence of X-ray outbursts and detection of X-ray pulsations, we adopt a log-uniform distribution in the range of $0.005-0.05$;
\item The X-ray luminosity $L_{\rm X,I}$ during type I outbursts:  we adopt a log-uniform distribution in the range of $10^{36}-10^{37}$ erg s$^{-1}$ \citep{Rei2011};
\item The X-ray luminosity $L_{\rm X,II}$ during type II outbursts:
\citet{Che2014} showed that the peak X-ray luminosity of the BeXPs is truncated around $2\times 10^{38}{\rm~erg~s^{-1}}$, and the distribution seems to prefer high-luminosity end. We accordingly adopt a uniform distribution in the range of $10^{37}-2\times10^{38}$ erg s$^{-1}$.
\item The X-ray luminosity in the quiescent phase $L_{\rm X,q}$: we adopt a log-uniform distribution in the range of $10^{34}-10^{36}$ erg s$^{-1}$ \citep{Yan2017};
\item The critical value of the fastness parameter $\omega_{\rm c}$: various studies show that it is in the range of $0.35-0.95$ \citep{Gho1979a,Gho1979b,Wan1995,Li1996,Lon2005,Zan2009,Zan2013}. In our work, we adopt a uniform distribution in the range of $0.5-1$;
\item The ADAF parameter $A$:  we adopt a uniform distribution in the the range $0.2-0.3$ \citep{Nar1995};
\end{itemize}
In next section we will vary the distribution forms and examine their influence.
Other (free) parameters are listed as follows,
\begin{itemize}
\item The disk wind fraction $\eta$;
\item The NS magnetic fields $B$: we adopt a log-normal distribution for $B$, i.e.
\begin{equation}
F({\rm lg} B ~|~ \mu_{{\rm lg}B},\sigma_{{\rm lg}B})=\frac{1}{\sqrt{2\pi \sigma_{{\rm lg}B}^2}}
{\rm exp}\left[-\frac{({\rm lg} B-\mu_{{\rm lg} B})^2}{2\sigma^2_{{\rm lg} B}}\right],
\end{equation}
where the average value $\mu_{{\rm lg} B}$ and the variance $\sigma_{{\rm lg} B}$ are left as free parameters;
\item The fraction $f$ of BeXPs that undergo type II outbursts;
\item The duty cycle $y$ of type II outbursts.
\end{itemize}

\begin{figure}
\gridline{\fig{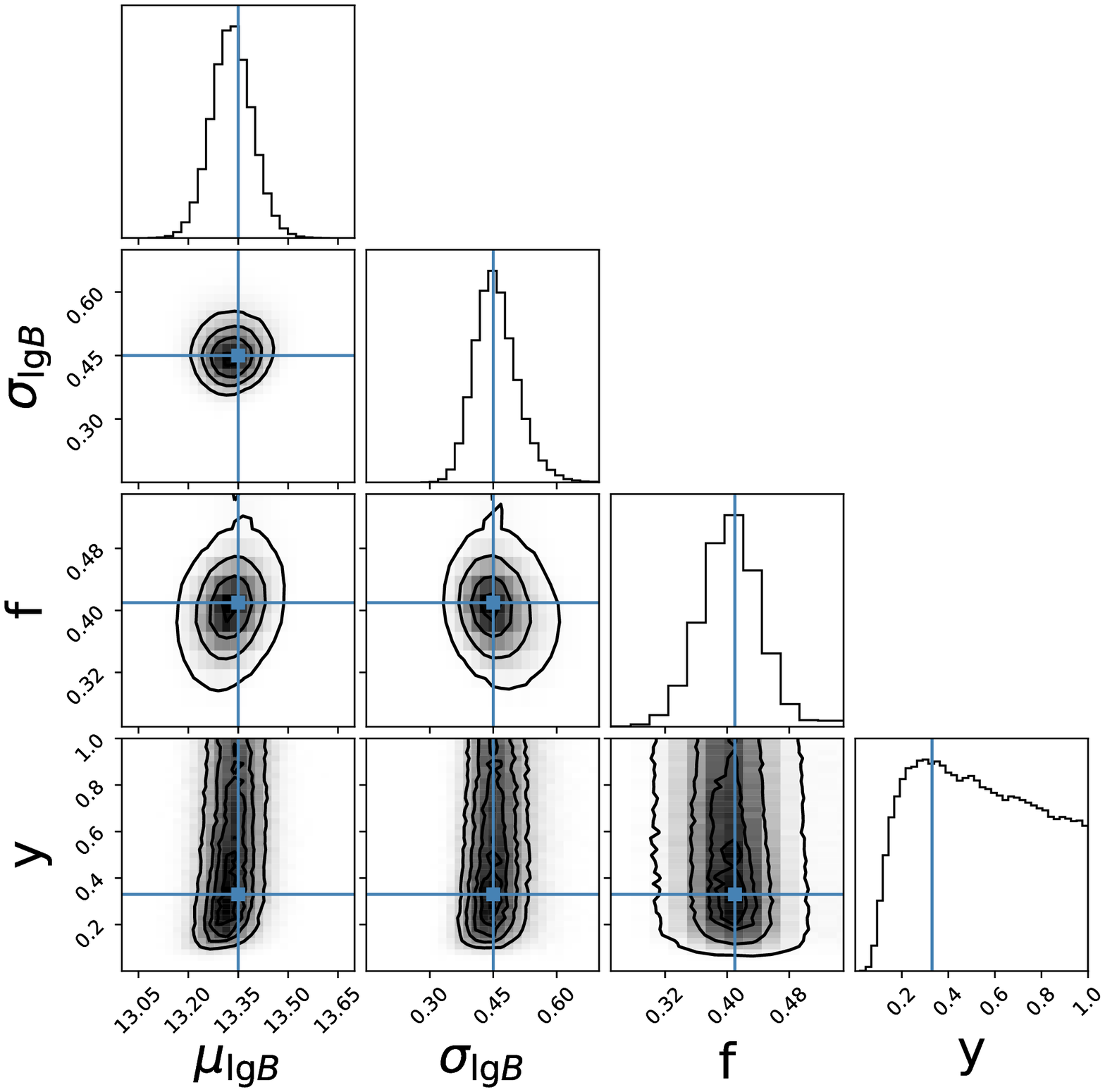}{0.3\textwidth}{(a1)}\fig{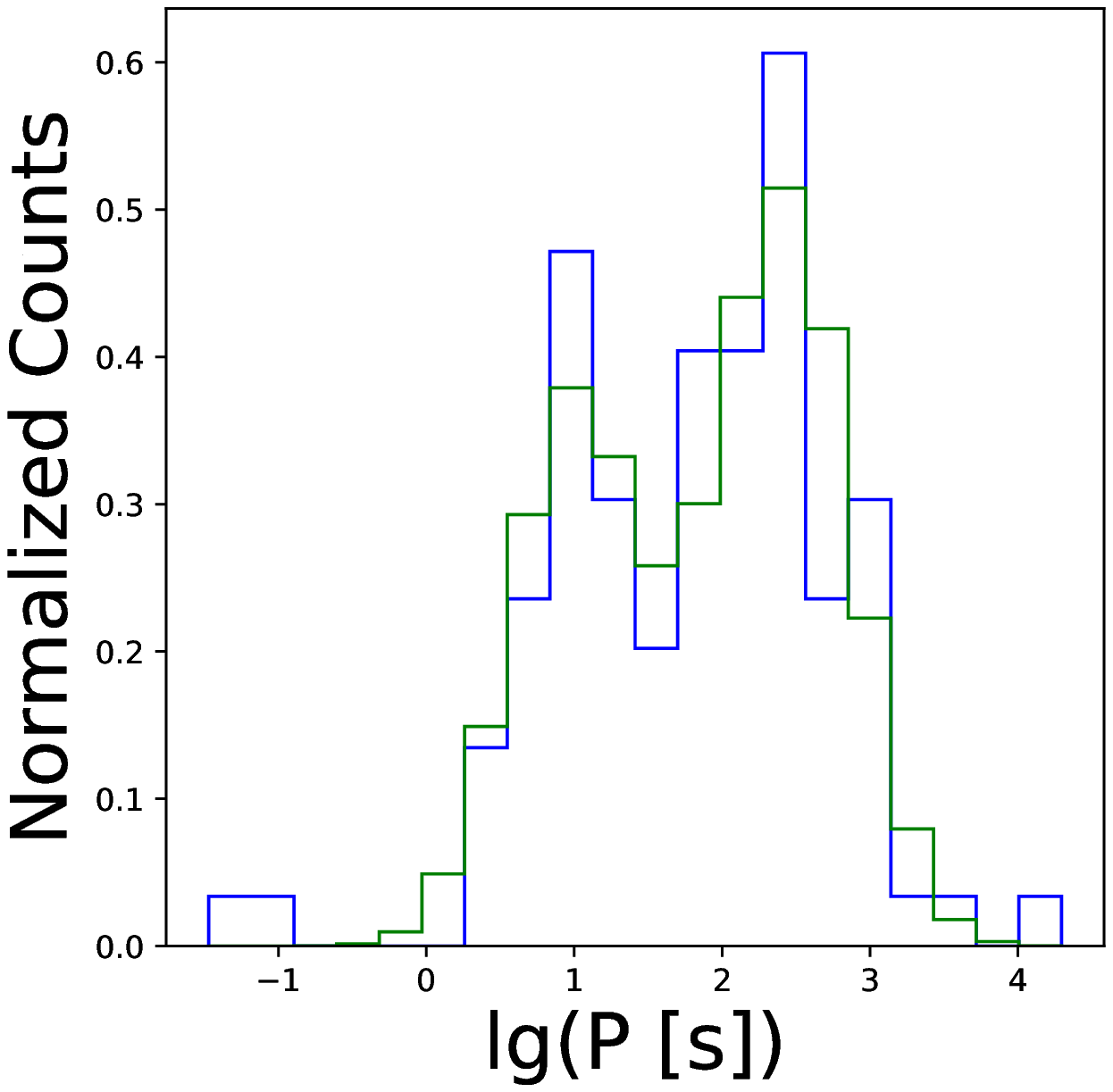}{0.3\textwidth}{(a2)}}
\gridline{\fig{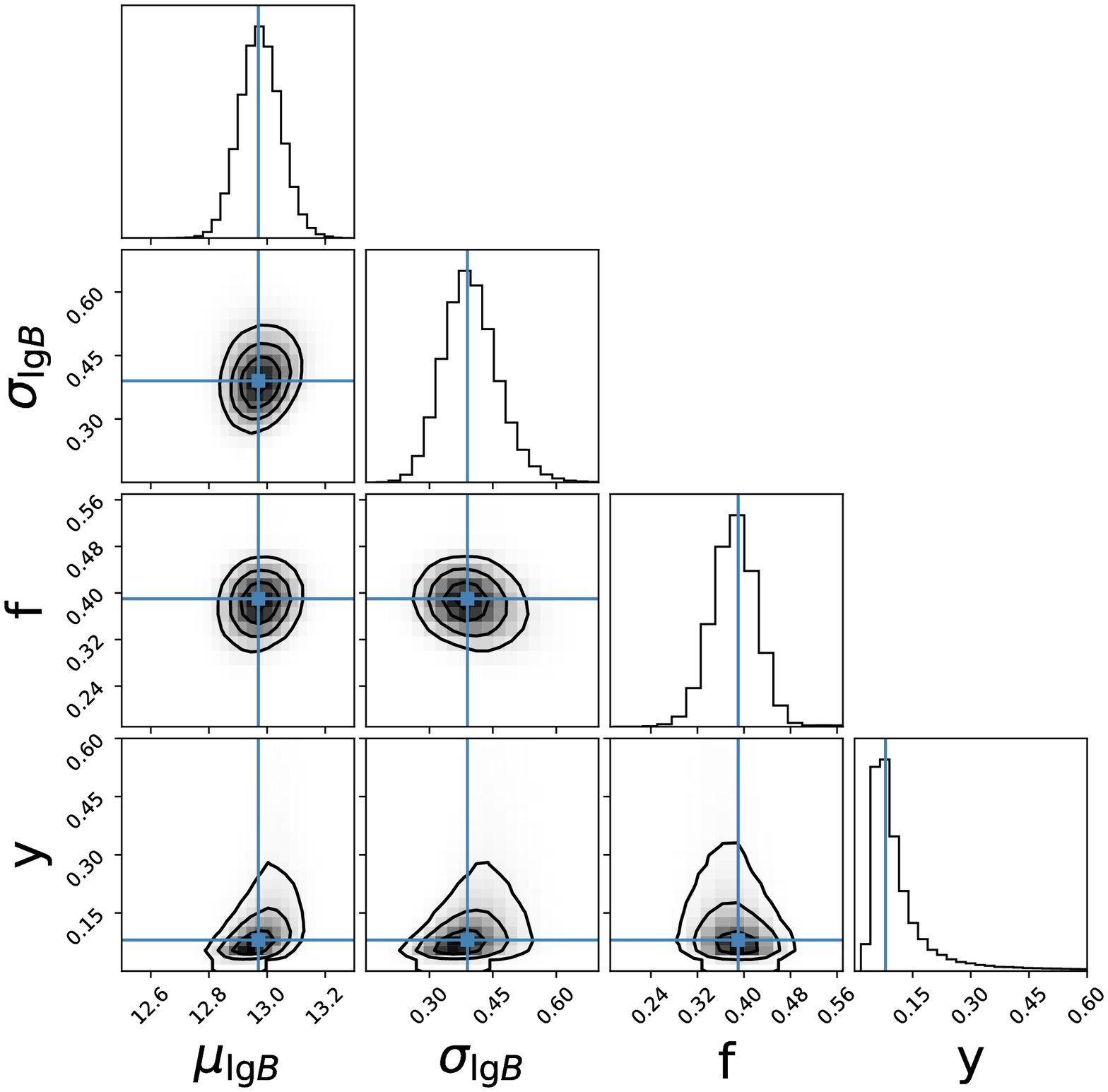}{0.3\textwidth}{(b1)}\fig{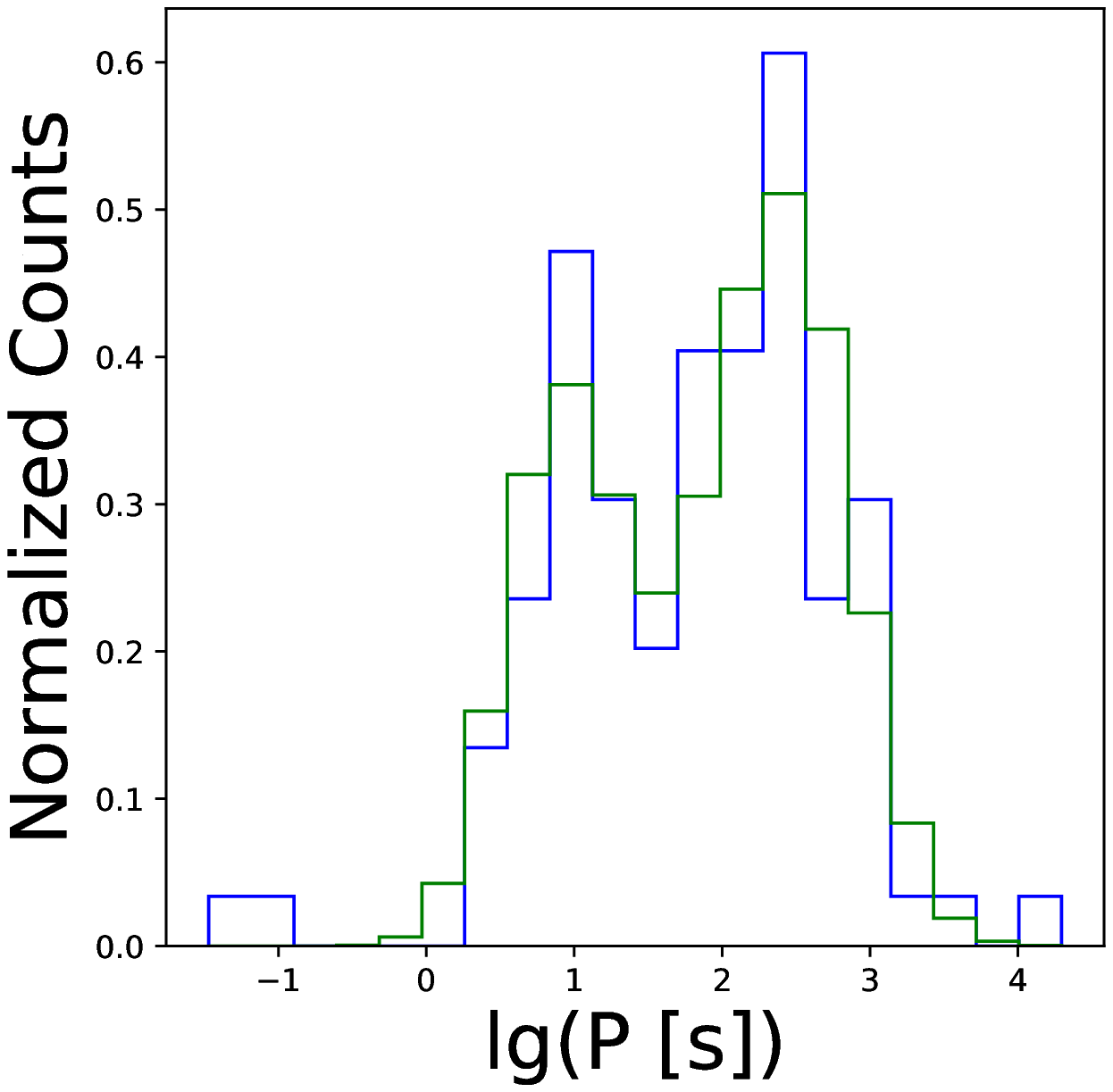}{0.3\textwidth}{(b2)}}
\gridline{\fig{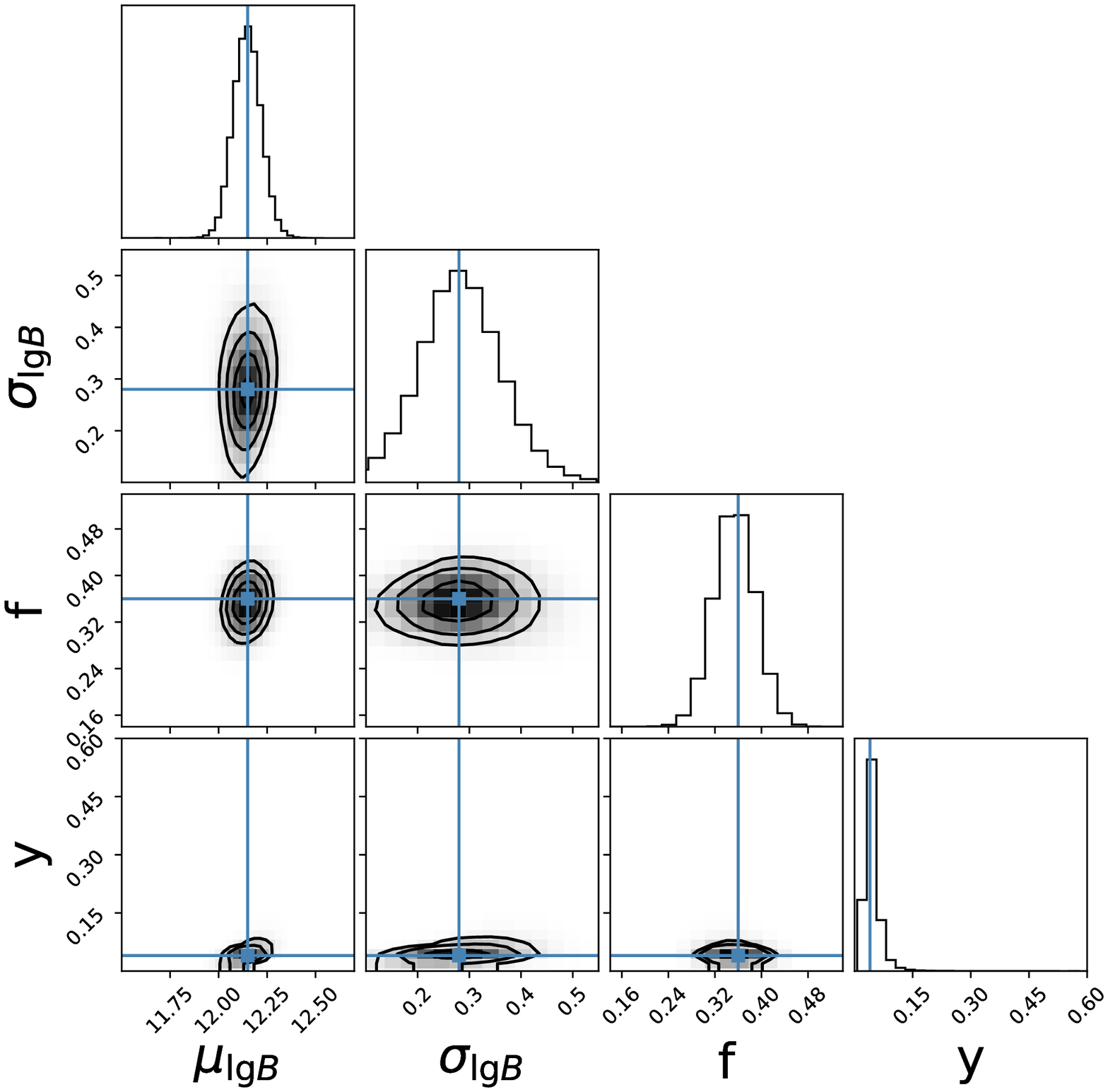}{0.3\textwidth}{(c1)}\fig{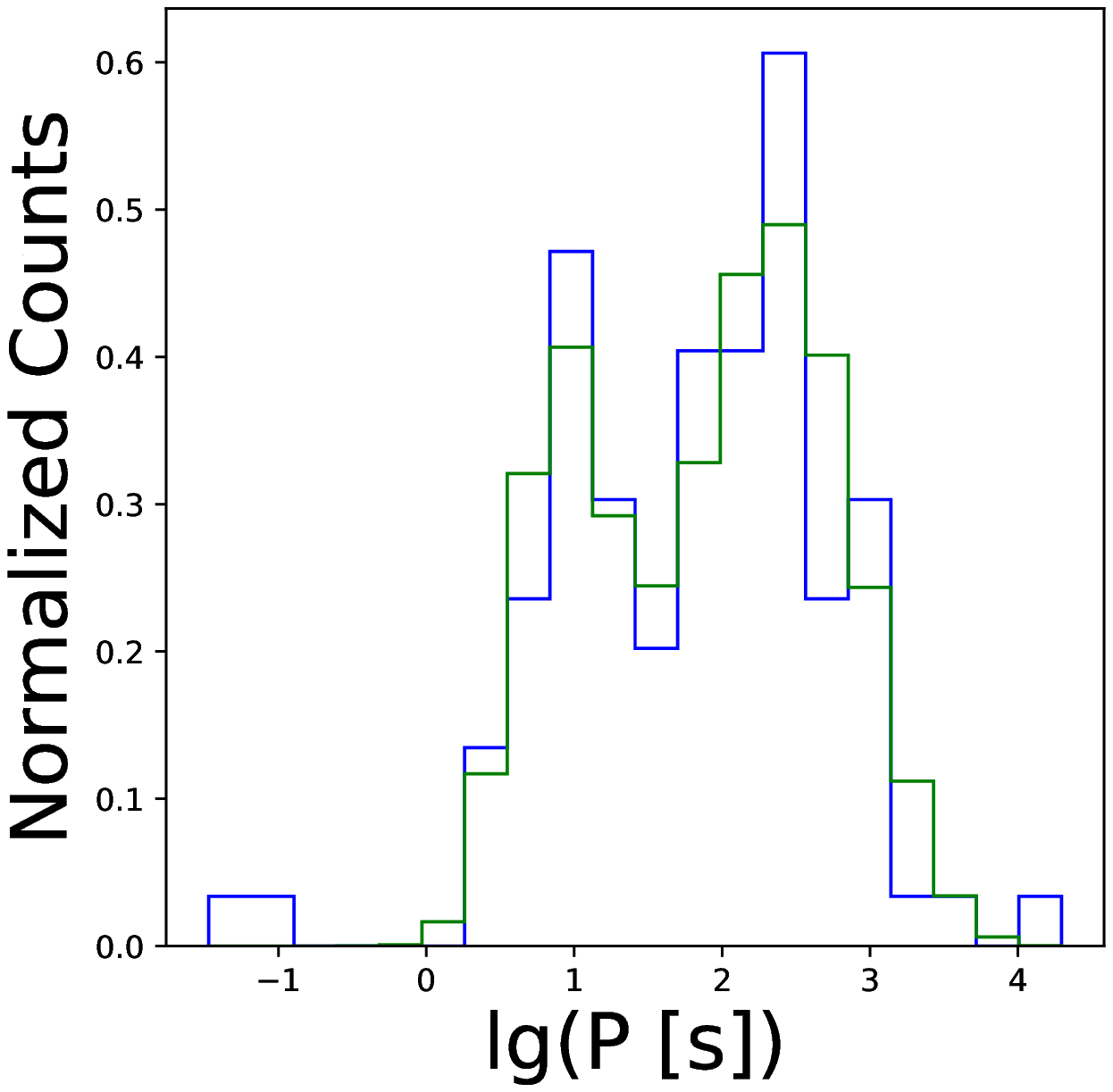}{0.3\textwidth}{(c2)}}
\caption{
The MCMC fitting.
The first, second, and third rows present the result of Solution-1, Solution-2, and Solution-3, respectively.
The left column shows the the one and two dimensional projections of the posterior probability distributions of $\mu_{\rm lg B}$, $\sigma_{\rm lg B}$, $f$, and $y$, where the blue solid lines represent the best fitting. The right column presents a comparison between the observed distribution (blue solid line) and the theoretical distribution (green solid line) given by the best fitting parameters.}
\label{f3}
\end{figure}

With the above parameters and the equilibrium spin period assumption we model the bimodal distribution of BeXPs with a Markov Chain Monte-Carlo method\footnote{The calculations were made with the emcee code, which is an open-source software for the Markov Chain Monte-Carlo method \citep{For2013}.}. We consider three types of solutions in our model: the no-wind solution with $\eta=0$, the weak-wind solution with $\eta=0.1$, and the strong-wind solution with $\eta=0.5$ (hereafter Solution-1, Solution-2, and Solution-3, respectively).
The calculated results are shown in Fig.~\ref{f3}
\footnote{The corner plots were generated by the corner.py module written by \citet{Dan2016}.}. There are three rows and two columns in each figure. The first, second, and third rows present the results of Solution-1, Solution-2, and Solution-3. The left column presents the posterior probability distribution of $\mu_{\rm lg B}$, $\sigma_{\rm lg B}$, $f$, and $y$, where the blue solid lines represent the best fitting points, and the right column presents the comparison between the observed distribution (blue solid lines) and the model distribution (green solid lines) with the best fitting parameters, which are summarized in Table \ref{tab}.

It is seen from Fig.~\ref{f3} and Table \ref{tab} that the three solutions can well reproduce the bimodal distribution of the spin periods. The short-period subpopulation mainly consists of BeXBs which undergo type II outbursts. There is a tendency that $\mu_{\rm lgB}$, $f$, and $y$ all become smaller with increasing $\eta$. This is not difficult to understand. As the wind torque acts to spin down the NS, a weaker magnetic field is required with stronger wind, given a specific equilibrium spin period.
When $\eta$ increases from 0 to 0.5, the average value of $\mu_{\rm lgB}$ decreases from 13.34 ($B\sim2.2\times 10^{13}$~G) to 12.15 ($B\sim1.4\times 10^{12}$~G), which are still in the reasonable range for X-ray pulsars.
The fraction $f$  of BeXBs with type II outbursts varies from 0.41 to 0.36.
According to the data in \citet{Che2014} we infer that the observed value of $f$ for transient BeXBs is roughly 0.2. But this should be taken as a lower limit since type II outbursts must have been missed because of limited monitoring by X-ray telescopes.
The duty cycle $y$ of type II outbursts changes from 0.33 to 0.04, which is roughly consistent with the observed value $\sim 0.1$.

We mention that the derived parameter distributions depend on the values of the input parameters which are subject to some uncertainties. For example, in our calculation, we use a constant luminosity during outbursts and ignore its evolution during the rising and decay phases for simplicity. The adopted distributions of $L_{\rm X,I}$, $L_{\rm X,II}$, and $L_{\rm X,q}$  are somewhat speculative and need to be refined by more dense observations. In addition, we neglect the possible influence of the eccentricities in BeXBs and the possible dependence of $f$ on both the orbital period and the eccentricity. To see how the specific distribution functions of the input parameters potentially affect our results, we  compare the results by taking into account different combinations of the distribution functions, with LU for log-uniform and U for uniform distributions. In the cases of Solution-2 and Solution-3, and we consider the following combinations for for $(x,~L_{\rm X,I},~L_{\rm X,II},~L_{\rm X,q})$,:  (LU, LU, U, LU), (U, LU, U, LU), (LU, U, U, LU, U), (LU, LU, U, U),  and (LU, LU, LU, LU). Our results are summarized in Table \ref{tab2}. We see that in these different combinations we obtain results that are compatible with each other. Therefore, our results are not sensitively dependent on the adopted distribution functions.

\section{Conclusions}

We summarize our work as follows.

(1) NSs in most BeXBs undergo spin-up during outbursts and spin-down during quiescence, so their spin evolution is controlled by the competition between the spin-up and spin-down torques. Therefore, it is inappropriate to use long-term, average steady accretion to calculate the spin evolution and derive the magnetic fields of the accreting NSs.

(2) Most NSs in BeXBs have reached the equilibrium spin periods since the spin-change timescale is sufficiently short compared with the ages of BeXBs.

(3) Considering the transient behavior of BeXBs, it seems able to reproduce the bimodal spin period distribution with the assumption of the equilibrium spin period for NSs with typical magnetic fields ($\sim 10^{12}-10^{13}$ G). The critical factors that determine the spin period distribution are the properties of type I and II outbursts.

\acknowledgments
This work was supported by the National Key Research and Development Program of China (2016YFA0400803), the Natural Science Foundation of China under grant Nos. 11333004, 11773015, 11573016, and Project U1838201 supported by NSFC and CAS.


\software{emcee \citep{For2013}, corner.py \citep{Dan2016}, scipy \citep{Eri2001}, R \citep{R2018}, diptest \citep{Mar2016}}
\appendix

\section{Detailed derivation of $P_{\rm eq}$ and $\tau_{\rm spin}$}
\begin{equation}
\begin{split}
I\frac{2\pi}{P^2}(-\dot{P})&=
\left[xN_{\rm 0,I}(1-\frac{\eta^{6/7}}{\phi^{1/2}})+
(1-x-y)N_{\rm 0,q}(A-\frac{\eta^{6/7}}{\phi^{1/2}})+
yN_{\rm 0,II}(1-\frac{\eta^{6/7}}{\phi^{1/2}})\right]\\
&-\left[xN_{\rm 0,I}\frac{\omega_{\rm I}}{\omega_{\rm c}}+(1-x-y)N_{\rm 0,q}\frac{\omega_{\rm q}}{\omega_{\rm c}}
+yN_{\rm 0,II}\frac{\omega_{\rm II}}{\omega_{\rm c}}\right]\\
&=
\frac{\phi^{1/2}(GM)^{3/7}\mu^{2/7}}{2^{1/14}}\left[x\dot{M}_{\rm I}^{6/7}(1-\frac{\eta^{6/7}}{\phi^{1/2}})+
(1-x-y)\dot{M}_{\rm q}^{6/7}(A-\frac{\eta^{6/7}}{\phi^{1/2}})+
y\dot{M}_{\rm II}^{6/7}(1-\frac{\eta^{6/7}}{\phi^{1/2}})\right]\\
&-\frac{2\pi \phi^2}{\omega_{\rm c}}\frac{\mu^{8/7}}{(2GM)^{2/7}}
\left[x\dot{M}_{\rm I}^{3/7}+(1-x-y)\dot{M}_{\rm q}^{3/7}+y\dot{M}_{\rm II}^{3/7}\right]P^{-1}.
\label{A1}
\end{split}
\end{equation}
Here we define two parameters $C_1$ and $C_2$, which are respectively given by
\begin{equation}
C_1=
\frac{\phi^{1/2}(GM)^{3/7}\mu^{2/7}}{2^{1/14}}\left[x\dot{M}_{\rm I}^{6/7}(1-\frac{\eta^{6/7}}{\phi^{1/2}})+
(1-x-y)\dot{M}_{\rm q}^{6/7}(A-\frac{\eta^{6/7}}{\phi^{1/2}})+
y\dot{M}_{\rm II}^{6/7}(1-\frac{\eta^{6/7}}{\phi^{1/2}})\right]\frac{1}{2\pi I}
\end{equation}
and
\begin{equation}
C_2=\frac{2\pi \phi^2}{\omega_{\rm c}}\frac{\mu^{8/7}}{(2GM)^{2/7}}
\left[x\dot{M}_{\rm I}^{3/7}+(1-x-y)\dot{M}_{\rm q}^{3/7}+y\dot{M}_{\rm II}^{3/7}\right]\frac{1}{2\pi I}.
\end{equation}
Then, Eq.~(\ref{A1}) can be rewritten as
\begin{equation}
\dot{P}=-(C_1P-C_2)P.
\label{A4}
\end{equation}
The equilibrium spin $P_{\rm eq}$ is given by setting $\dot{P}=0$, i.e.,
\begin{equation}
P_{\rm eq}=\frac{C_2}{C_1}.
\label{A5}
\end{equation}
Assuming that the torque generated by the wind is smaller than the material toque, mathematically $(1-\eta^{6/7}/\phi^{1/2}) > 0$ and $(A-\eta^{6/7}/\phi^{1/2})>0$, we have $C_1 \neq0$. Then, Eq. (\ref{A4}) has following solution
\begin{equation}
P=\frac{C_2/C_1}{1-(1-C_2/C_1/P_0)e^{-C_2T}},
\end{equation}
where $T$ is the evolutionary time, and $P_0$ is the spin period at $T=0$. Note that, according to Eq.~(\ref{A5}), $C_2/C_1$ can be substituted by $P_{\rm eq}$. Then, the spin evolution is given by
\begin{equation}
P=\frac{P_{\rm eq}}{1-(1-P_{\rm eq}/P_0)e^{-C_2T}}.
\label{A7}
\end{equation}
According to the exponential term in Eq.~(\ref{A7}), we can define the spin-evolutionary timescale $\tau_{\rm spin}$,
\begin{equation}
\tau_{\rm spin}=C_2^{-1}.
\end{equation}

\bibliographystyle{aasjournal}

\newpage

\begin{deluxetable*}{c|cccc}[]
\tablecaption{Fitting parameters for the period distribution$^{\rm a}$ \label{tab}}
\tablecolumns{6}
\tablenum{1}
\tablewidth{0pt}
\tablehead{
\colhead{$\eta$} &
\colhead{$\mu_{{\rm lg} B}$} &
\colhead{$\sigma_{{\rm lg} B}$} &
\colhead{$f$} &
 \colhead{$y$} }
\startdata
 0 & 13.34$_{-0.11}^{+0.09}$ & 0.45$_{-0.07}^{+0.10}$ & 0.41$_{-0.06}^{+0.40}$   & 0.33$_{-0.17}^{+0.62}$ \\
 0.1& 12.97$_{-0.11}^{+0.12}$ & 0.39$_{-0.09}^{+0.12}$  & 0.39$_{-0.07}^{+0.07}$  & 0.08$_{-0.04}^{+0.34}$\\
 0.5& 12.15$_{-0.12}^{+0.12}$ & 0.28$_{-0.09}^{+0.12}$  & 0.36$_{-0.06}^{+0.05}$ & 0.04$_{-0.01}^{+0.04}$\\
\enddata
\tablenotetext{a}{The errors are given under 90\% confidence level.}
\end{deluxetable*}

\begin{deluxetable*}{c|cccc|cccc}[b!]
\tablecaption{Fitting parameters for different combinations of the distribution functions \label{tab2}}
\tablecolumns{9}
\tablenum{2}
\tablewidth{0pt}
\tablehead{
\colhead{$\eta$} &
\multicolumn{4}{c}{distributions of parameters\tablenotemark{a}}&
\colhead{$\mu_{{\rm lg} B}$} &
\colhead{$\sigma_{{\rm lg} B}$} &
\colhead{$f$} &
 \colhead{$y$} \\
\cline{2-5}
 \colhead{}& \colhead{$x$} & \colhead{$L_{\rm X,I}$} & \colhead{$L_{\rm X,II}$}& \colhead{$L_{\rm X,q}$} & \colhead{} & \colhead{} & \colhead{}
 }
\startdata
{}   & LU & LU & U  &LU & 12.97$_{-0.11}^{+0.12}$ & 0.39$_{-0.09}^{+0.12}$  & 0.39$_{-0.07}^{+0.07}$  & 0.08$_{-0.04}^{+0.34}$\\
{}       & U & LU & U & LU&13.12$_{-0.10}^{+0.11}$ & 0.40$_{-0.08}^{+0.10}$  & 0.37$_{-0.05}^{+0.08}$ & 0.15$_{-0.04}^{+0.77}$\\
0.1       & LU & U & U & LU&13.10$_{-0.10}^{+0.11}$ & 0.40$_{-0.09}^{+0.10}$  & 0.37$_{-0.05}^{+0.07}$  & 0.15$_{-0.06}^{+0.74}$\\
{}       & LU & LU & U & U  & 12.87$_{-0.11}^{+0.11}$ & 0.41$_{-0.08}^{+0.12}$  & 0.40$_{-0.06}^{+0.41}$ & 0.07$_{-0.02}^{+0.09}$\\
{}       & LU & LU & LU &LU  & 12.96$_{-0.09}^{+0.11}$ & 0.38$_{-0.08}^{+0.12}$  & 0.37$_{-0.05}^{+0.08}$ & 0.18$_{-0.06}^{+0.72}$\\
\hline
{}   & LU & LU & U  &LU & 12.15$_{-0.12}^{+0.12}$ & 0.28$_{-0.09}^{+0.12}$  & 0.36$_{-0.06}^{+0.05}$ & 0.04$_{-0.01}^{+0.04}$\\
{}       & U & LU & U & LU&12.39$_{-0.14}^{+0.10}$ & 0.35$_{-0.11}^{+0.11}$  & 0.36$_{-0.06}^{+0.06}$ & 0.08$_{-0.02}^{+0.57}$\\
0.5       & LU & U & U & LU&12.35$_{-0.13}^{+0.11}$ & 0.32$_{-0.10}^{+0.13}$  & 0.36$_{-0.06}^{+0.05}$  & 0.06$_{-0.01}^{+0.43}$\\
{}       & LU & LU & U & U  & 11.90$_{-0.15}^{+0.08}$ & 0.28$_{-0.12}^{+0.15}$  & 0.37$_{-0.07}^{+0.05}$ & 0.03$_{-0.01}^{+0.02}$\\
{}       & LU & LU & LU &LU  & 12.15$_{-0.11}^{+0.12}$ & 0.27$_{-0.17}^{+0.14}$  & 0.36$_{-0.06}^{+0.05}$ & 0.09$_{-0.03}^{+0.26}$\\
\enddata
\tablenotetext{a}{LU and U for log-uniform distribution and uniform distribution respectively.}
\end{deluxetable*}

\newpage

\end{document}